\documentclass[aps,preprint,floatfix,prb,showpacs,amsmath,amssymb]{revtex4}

\usepackage{graphicx,color}
\usepackage{dcolumn}
\newcolumntype{.}{D{.}{.}{-1}}
\usepackage{bm}
\usepackage[latin1]{inputenc}
\usepackage{epstopdf}
\usepackage{subfigure}

\newcommand{\tc}{$T_{\rm C}$}
\newcommand{\peff}{$p_{\rm eff}$}
\newcommand{\geff}{$g_{\rm eff}$}
\newcommand{\mb}{\(\mu _{\rm B}\)}

\newcommand{\tn}{$T_{\rm N}$}
\newcommand{\mn}{MnSb$_2$O$_6$}
\newcommand{\cu}{CuSb$_2$O$_6$}
\newcommand{\bacu}{BaCu$_2$Si$_2$O$_7$}
\newcommand{\bif}{BiFeO$_3$}
\newcommand{\life}{LiFePO$_4$}
\newcommand{\etal}{\textit{et~al.}}

\newcommand{\cpp}{$c_{\rm p}^{\rm ph}$}
\newcommand{\cp}{$c_{\rm p}$}
\newcommand{\dcp}{$\Delta c_{\rm p}$}

\newcommand{\DS}{$\Delta S^{\rm mag}$}
\newcommand{\jmk}{J/(mol$\cdot$K)}

\newcommand{\bc}{$B_{\rm C}$}
\newcommand{\db}{$\Delta B$}
\newcommand{\bco}{$B_{\rm C1}$}
\newcommand{\bct}{$B_{\rm C2}$}
\newcommand{\bmax} {$B_{\text{res}}^{\text{max}}$}
\newcommand{\bmin} {$B_{\text{res}}^{\text{min}}$}
\newcommand{\bres} {$B_{\text{res}}$}

\begin{document}

\title{Magnetic anisotropy and the phase diagram of chiral MnSb$_2$O$_6$ }


\author{J.~Werner}
\affiliation{Kirchhoff Institute of Physics, Heidelberg University, INF 227, D-69120 Heidelberg, Germany}
\affiliation{Centre for Advanced Materials, Heidelberg University, INF 225, 69120 Heidelberg, Germany}
\author{C.~Koo}
\affiliation{Kirchhoff Institute of Physics, Heidelberg University, INF 227, D-69120 Heidelberg, Germany}
\affiliation{Centre for Advanced Materials, Heidelberg University, INF 225, 69120 Heidelberg, Germany}
\author{R.~Klingeler}
\email[Email:]{klingeler@kip.uni-hd.de}
\affiliation{Kirchhoff Institute of Physics, Heidelberg University, INF 227, D-69120 Heidelberg, Germany}
\affiliation{Centre for Advanced Materials, Heidelberg University, INF 225, 69120 Heidelberg, Germany}
\author{A.N.~Vasiliev}
\affiliation{Physics Faculty, M.V. Lomonosov Moscow State University, Moscow 119991, Russia}
\affiliation{Theoretical Physics and Applied Mathematics Department, Ural Federal University, 620002 Ekaterinburg, Russia}
\affiliation{National University of Science and Technology "MISiS", Moscow 119049, Russia}
\author{Y.A.~Ovchenkov}
\affiliation{Physics Faculty, M.V. Lomonosov Moscow State University, Moscow 119991, Russia}
\author{A.S.~Polovkova}
\affiliation{Physics Faculty, M.V. Lomonosov Moscow State University, Moscow 119991, Russia}
\author{G.V.~Raganyan}
\affiliation{Physics Faculty, M.V. Lomonosov Moscow State University, Moscow 119991, Russia}
\author{E.A.~Zvereva}
\affiliation{Physics Faculty, M.V. Lomonosov Moscow State University, Moscow 119991, Russia}


\date{\today}

\begin{abstract}
The magnetic phase diagram and low-energy magnon excitations of structurally and magnetically chiral \mn\ are reported. The specific heat and the static magnetization are investigated in magnetic fields up to 9~T and 30~T, rspectively, while the dynamic magnetic properties are probed by X-band as well as tunable high-frequency electron spin resonance spectroscopy. Below \tn\ = 11.5~K, we observe antiferromagnetic resonance modes which imply small but finite planar anisotropy showing up in a zero-field splitting of 20~GHz. The data are well described by means of an easy-plane two-sublattice model with the anisotropy field $B_A=0.02$~T. The exchange field $B_{\rm E}= 13$~T is obtained from the saturation field derived from the pulsed-field magnetisation. A crucial role of the small anisotropy for the spin structure is reflected by competing antiferromagnetic phases appearing, at $T = 2$~K, in small magnetic fields at \bco\ $\approx$ 0.5~T and \bct\ = 0.9~T. We discuss the results in terms of spin reorientation and of small magnetic fields favoring helical spin structure over the cycloidal ground state which, at $B=0$, is stabilized by the planar anisotropy. Above \tn , short-range magnetic correlations up to $\gtrsim 60$~K and magnetic entropy changes well above \tn\ reflect the frustrated triangular arrangement of Mn$^{2+}$-ions in \mn .
\end{abstract}

\maketitle

\section{Introduction}

Frustrated or low-dimensional magnetic materials exhibiting low-energy magnetic excitations in which complex magnetic order is coupled to structure and dielectric properties are promising candidates for multiferroic properties.~\cite{lottermoser,eerenstein, cheongreview, fiebigreview} On the microscopic point of view, magnetic anisotropy has been proven an essential ingredient in a multitude of systems with significant magnet-electric coupling. In \bif , where both magnetic ordering at \tn\ = 650~K and ferroelectricity at \tc\ = 1100~K appear well above room temperature~\cite{bif1}, anisotropy is crucial for stabilising the actual cycloidal spin configuration.~\cite{Matsuda2012} This also holds for \mn , a structurally and magnetically chiral system which is predicted to be multiferroic with a unique ferroelectric switching mechanism based on its corotating cycloidal magnetic structure.~\cite{Johnson13} Despite its relevance, the size of magnetic anisotropy has not been experimentally determined yet.

\mn\ crystallizes in the trigonal space group P321 and exhibits structural chirality.~\cite{Reimers,Johnson13} In the layered structure, triples of MnO$_6$ distorted octahedra connected by SbO$_6$ octahedra, thereby forming isolated triangles of magnetic Mn$^{2+}$ ions. Below \tn\ = 12.5~K, incommensurate long range antiferromagnetic order evolves with the spin structure being based  on corotating cycloids.~\cite{Reimers} The experimentally observed magnetic structure can be derived from ab-initio density functional theory (DFT) calculations with magnetic anisotropy playing a crucial role for establishing the ground state.~\cite{Johnson13} We hence present a detailed study of the static and dynamic magnetic properties of \mn\ with particular emphasis on magnetic anisotropy. Experimentally, we apply specific heat and magnetisation measurements in static and pulsed magnetic field as well as X-band (LF-ESR) and tunable high-frequency/high-field electron spins resonance (HF-ESR) studies. The magnetic phase diagram of \mn\ shows three antiferromagnetic phases, two of which limited to small magnetic fields $\simeq 1$~T, i.e. well below the saturation field of about 26~T. The antiferromagnetic resonance (AFMR) modes probed by HF-ESR imply an easy plane-type behavior and a zero field splitting of approximately 20~GHz. Broadening of the HF-ESR resonances and magnetic entropy changes well above \tn\ suggest short range AFM fluctuations up to $\gtrsim 60$~K.

\section{Experimental}

Polycrystalline \mn\ was prepared by conventional solid state synthesis as reported elsewhere.~\cite{Nalbandyan2015} Phase purity was confirmed by X-ray diffraction. Refined hexagonal lattice parameters ($a=8.8035(17)$, $c=4.7266(12)$~\AA) are in a good agreement with the literature.~\cite{Scott89,Reimers}

HF-ESR measurements were carried out using a phase-sensitive millimeter-wave vector network analyzer (MVNA) from AB Millimetr\'{e} covering the frequency range from 30 to 1000 GHz.~\cite{Comba2015} For each frequency range (Q,L,W band etc.), different sets of Schottky diode systems were used. Experiments were performed in a 18 T superconducting magnet with temperature control sensors in both probe and sample space. \mn\ loose powder was placed in the sample space of the cylindrical waveguide probe without any glue or grease. LF-ESR studies were carried out in an X-band ESR spectrometer CMS 8400 (ADANI) ($f\approx 9.4$~GHz, $B\leq 0.7$~T) with a BDPA (a,g-bisdiphenyline-b-phenylallyl) reference sample which labels $g = 2.00359$. Static magnetic properties were measured with a Quantum Design MPMS XL-5 SQUID magnetometer and a Quantum Design PPMS-9 system, respectively. The latter system has been applied to obtain the specific heat, too. Magnetisation studies in pulsed magnetic fields up to 30~T were performed in a magnet with a rise time of about 8~ms.

\section{Static magnetisation and magnetic phase diagram}

\begin{figure}
\includegraphics[width=1.0\columnwidth,clip] {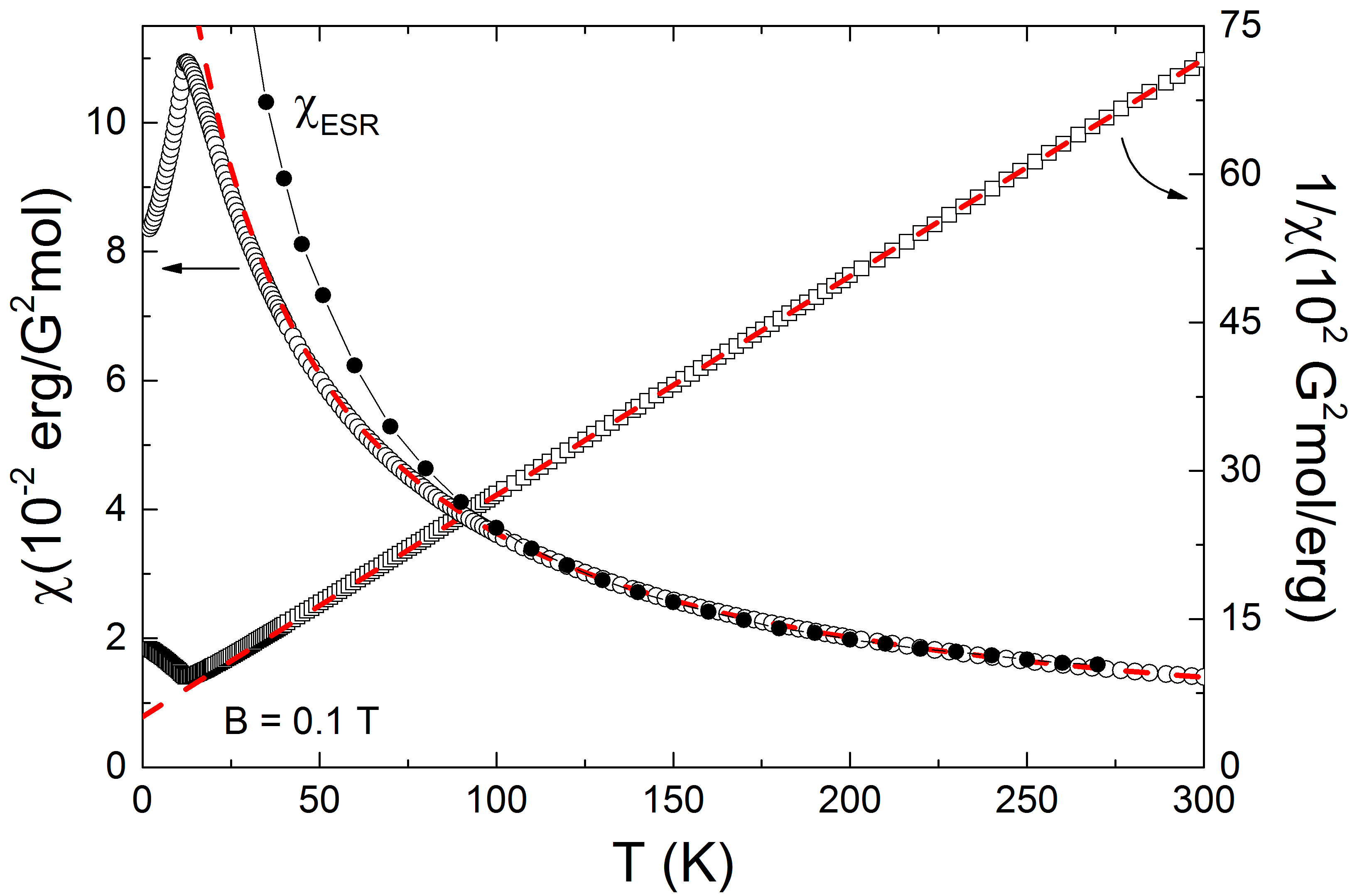}
\caption{Static magnetic susceptibility $\chi = M/B$ of \mn , at B = 0.1 T, and $\chi_{\rm ESR}$ obtained from doubly integrating the LF-ESR spectra, vs. temperature. The dashed lines display a Curie-Weiss approximation to the high temperature regime with \peff\ = 5.93 \mb\ and $\Theta = -23$ K (see the text).}\label{chi}
\end{figure}

\begin{figure}
\includegraphics[width=1.0\columnwidth,clip] {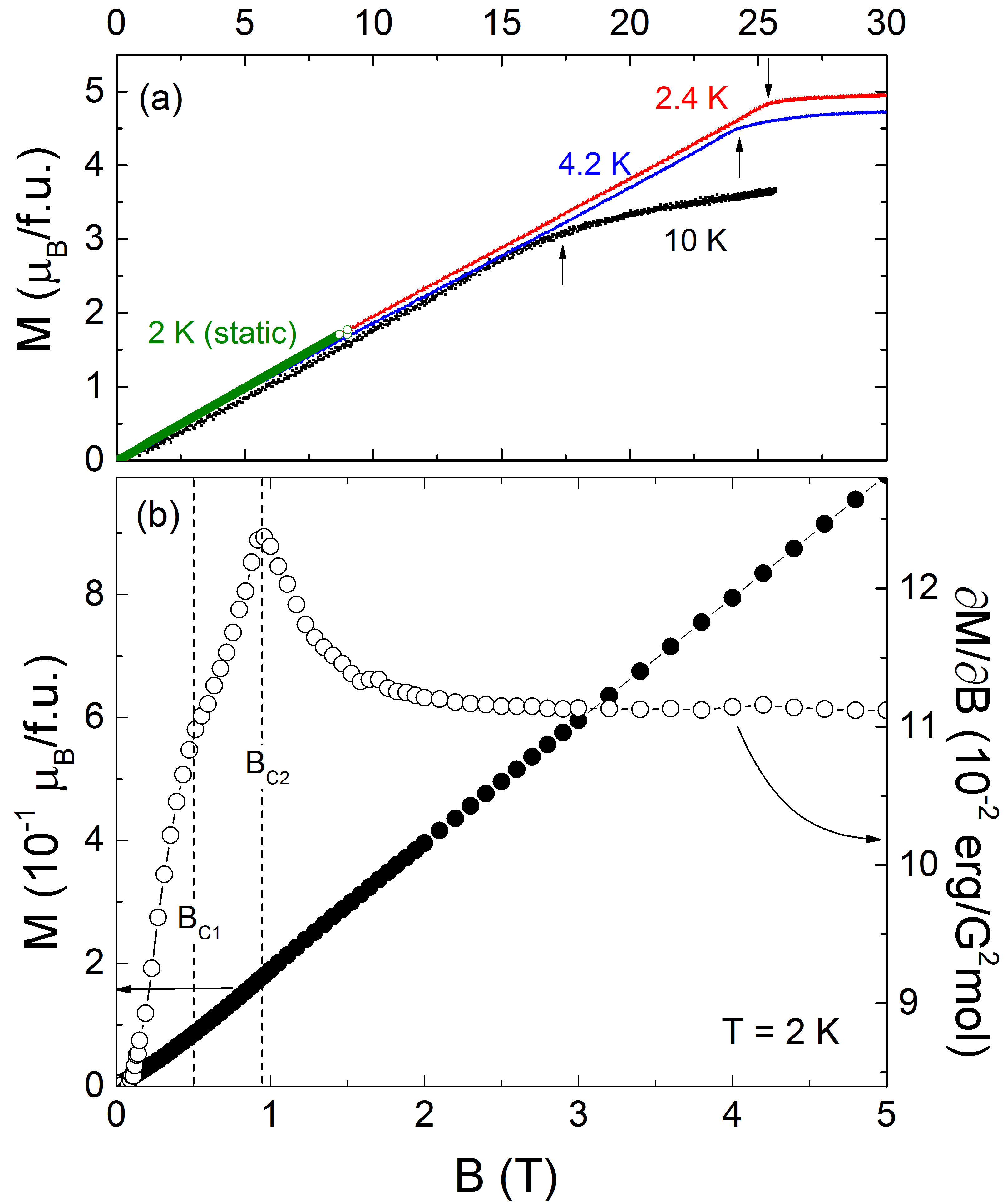}
\caption{(a) Pulsed field magnetisation vs. external field at $T$ = 2.4, 4.2, and 10~K are shown together with data (open circles) obtained in quasi-static field. The arrows mark the saturation fields \bc , \textsl{viz.}, the antiferromagnetic phase boundary. (b) Magnetisation $M$ and its derivative $\partial M/\partial B$ of \mn\ vs. external magnetic field, at $T=2$ K. The dashed lines indicate two anomalies at \bco\ and \bct .} \label{MB}
\end{figure}

\begin{figure}
\includegraphics[width=1.0\columnwidth,clip] {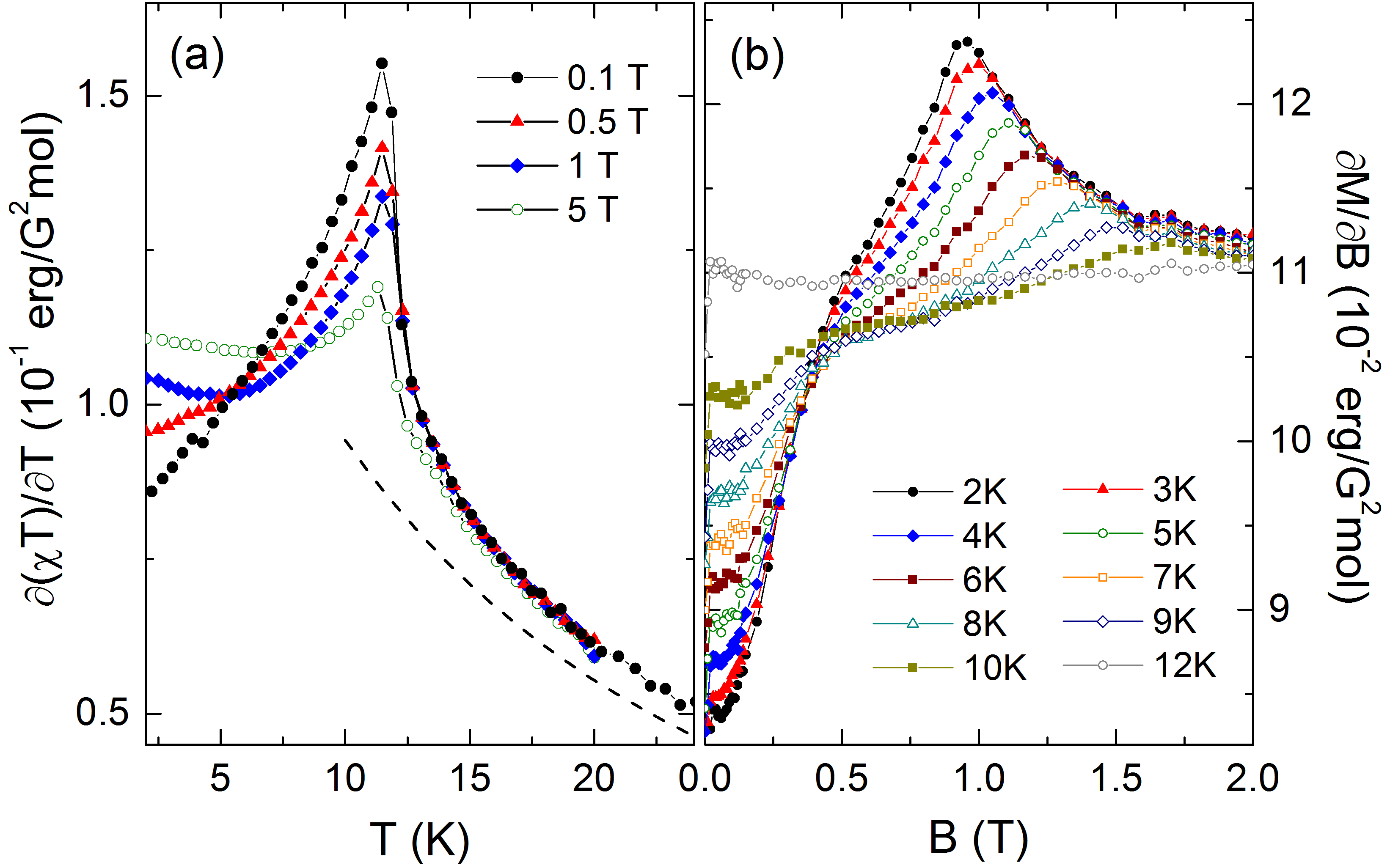}
\caption{(a) Magnetic specific heat $c_{\rm p}^{\rm magn}=\partial(\chi\cdot T)/\partial T$ as derived from the static susceptibility, indicating  a $\lambda$-like anomaly at \tn\ = 12 K, at different external fields. The dashed line shows the Curie-Weiss behavior $\partial(\chi_{\rm cw}\cdot T)/\partial T$. (b) Derivative $\partial M/\partial B$ of the magnetisation vs. external magnetic field, at different temperatures.} \label{MB2}
\end{figure}

The static magnetic susceptibility $\chi$ = $M$/$B$ of \mn\ (Fig.~\ref{chi}) confirms long range antiferromagnetic order below \tn\ = 11.5(5)~K. The onset of long range magnetic order is particularly evident in the magnetic specific heat $c_{\rm p}^{\rm magn}=\partial(\chi\cdot T)/\partial T$ which exhibits a pronounced $\lambda$-like anomaly at \tn\ (Fig.~\ref{MB2}a).~\cite{Fisher} Note, that in Ref.~[\onlinecite{Reimers}] a slightly different value of \tn\ was derived from the maximum in $\chi$. In the high temperature regime $\chi$ obeys a Curie-Weiss law $\chi_{\rm cw} = \chi_0+(N_{\rm A}\cdot p_{\rm eff}^2)/(3k_B (T+\Theta))$, with $\chi_0$ being a temperature independent contribution, $N_{\rm A}$ the Avogadro number, $p_{\rm eff}$ the effective magnetic moment, $k_{\rm B}$ the Boltzmann constant, and $\Theta$ the Weiss temperature. Fitting the data by means of the Curie-Weiss equation yields \peff\ = 5.93(2)~\mb , $\Theta = -23(1)$~K, and $\chi_0 = 2(1)\cdot 10^{-4}$~erg/(G$^2\cdot$mol). The obtained effective moment nearly perfectly agrees to what is expected for high-spin Mn$^{2+}$-ions with $S=5/2$ and $g = 1.995$, the latter found in our ESR measurements (see below). Below $T\approx 55$~K, the mean-field description starts to deviate from the experimental data indicating the onset of antiferromagnetic fluctuations, while long-range antiferromagnetic spin order evolves at \tn .

\begin{figure}[t]
\includegraphics[width=1.0\columnwidth] {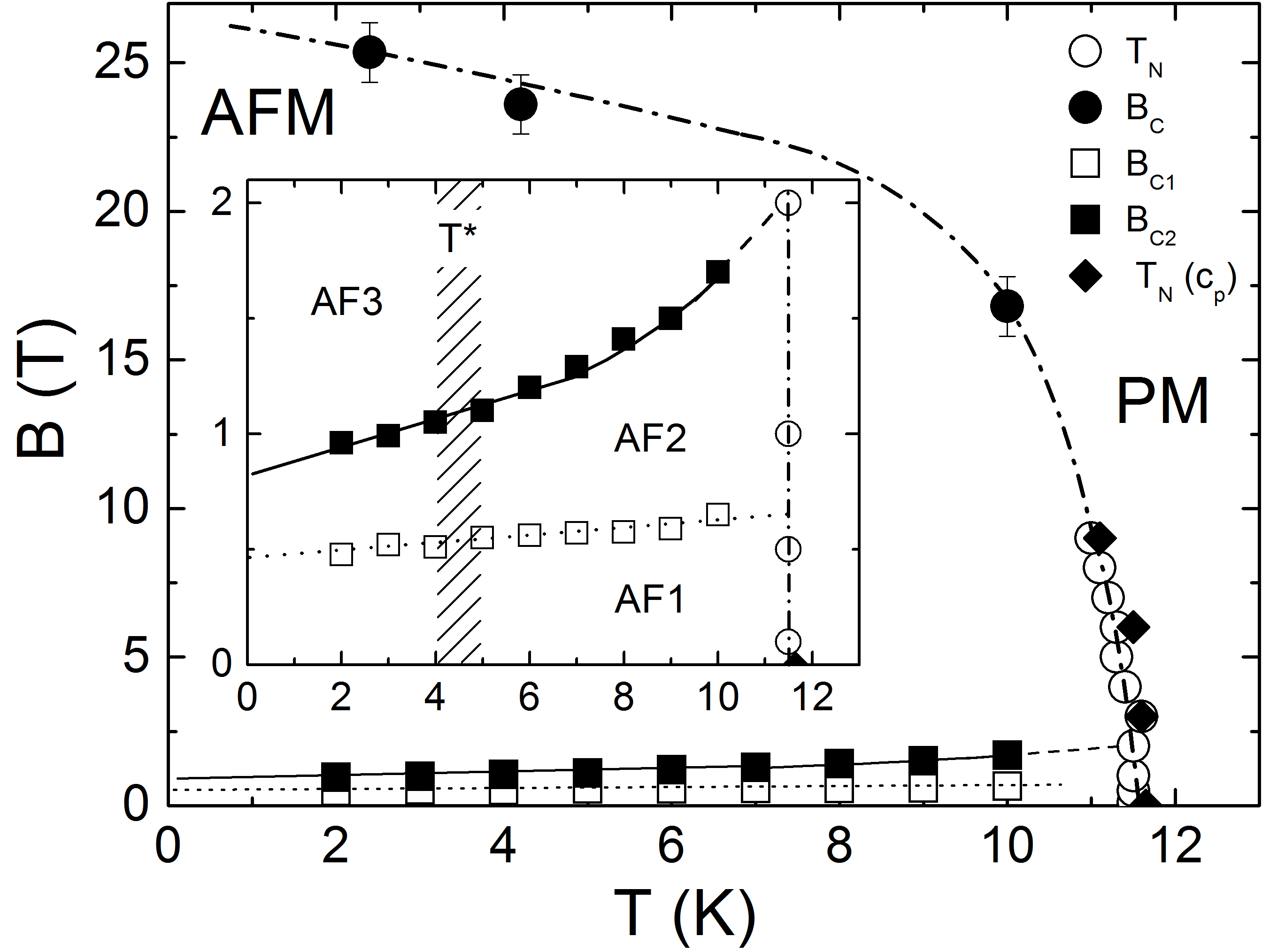}
\caption{Magnetic phase diagram of \mn . The dotted-dashed line shows the magnetic field dependence of \tn\ as derived from $M$ and \cp\ vs. $T$ measurements at different magnetic fields (cf. Figs.~\ref{MB2}a and \ref{cp}). The dotted and straight lines refer to \bco\ and \bct , respectively, taken from $M$ vs. $B$ (cf. Fig.~\ref{MB2}b and Fig.~\ref{MB}). AF1, AF2, and AF3 label the different antiferromagnetic phases, PM the paramagnetic one. The shaded area reflects the hump-like feature observed in the specific heat data at $T$*.} \label{phd}
\end{figure}

\begin{figure}[t]
\includegraphics[width=1.0\columnwidth] {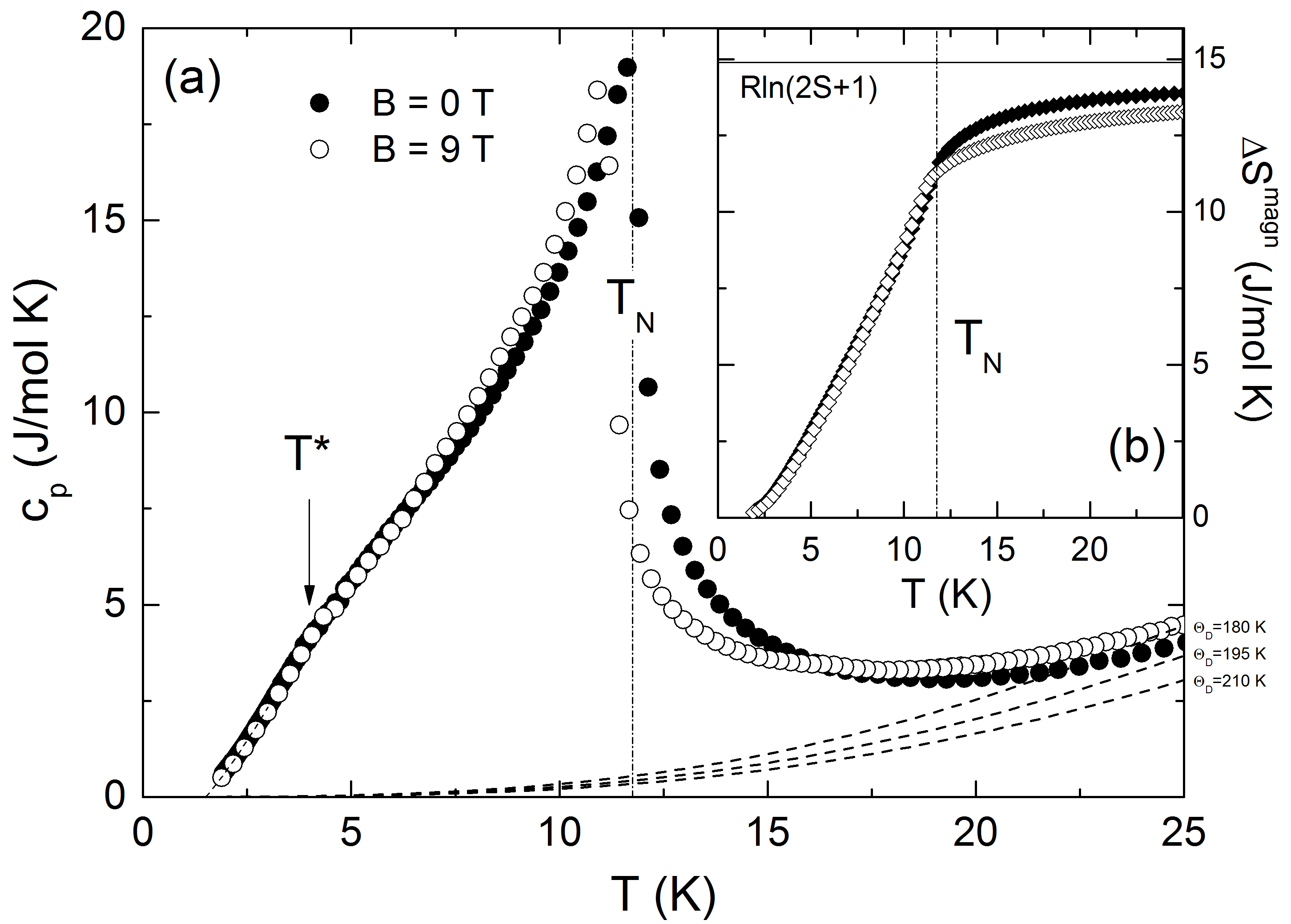}
\caption{(a) Specific heat \cp\ and (b) magnetic entropy changes of \mn . The dashed lines show the phonon contribution \cpp\ according to $\Theta_D = 180$, 195, and 210~K, respectively. Dashed-dotted lines show \tn\ at $B=0$ and the dotted line extrapolates \cp\ to low temperatures. Entropy changes in (b) have been calculated by integrating (\cp -\cpp )/$T$.} \label{cp}
\end{figure}

The field dependence of the magnetisation $M(B)$ in Fig.~\ref{MB} corroborates antiferromagnetic behaviour at low temperatures. For $B>2$~T, there is a linear field dependence of $M$. At high fields, right bending of the magnetisation curves obtained in pulsed magnetic fields up to $B=30$~T shown in Fig.~\ref{MB}a indicates the antiferromagnetic phase boundary (cf. Fig.~\ref{phd}). At $T=2.4$~K, the associated critical field is \bc =25.3~T and the observed value $M$($T$=2.4~K,$B$=25.3~T) = 4.9~\mb /f.u. is close to the theoretical saturation magnetization. Extrapolating the data suggests the saturation field at zero temperature of \bc\ $\approx 26$~T. At small magnetic fields, the $M$ vs. $B$ curve is slightly left-bending which is often associated to magnetic anisotropy. In contrast to a typical spin flop-like behavior, in \mn\ left-bending is associated with two separated anomalies at \bco\ $\approx$ 0.5~T and \bct\ = 0.9~T, at $T=2$~K. As seen in Fig.~\ref{MB2}b, both anomalies are restricted to the long range antiferromagnetic ordered phase. Upon heating from $T=2$~K, \bco\ is essentially constant while \bct\ clearly increases and significantly broadens. E.g., at 9~K, \bct\ amounts to 1.5~T. The field dependence of \tn\ is derived from the specific heat shown in Figs.~\ref{MB2} and \ref{cp}. At low temperatures, the data clearly deviate from the Curie-Weiss behaviour but obey the mean-field description at $T \gtrsim 55$~K.

The observed positive slope of the phase boundary \bct ($T$) agrees to the associated increase in magnetisation. Quantitatively, the slope of the phase boundary is associated with the ratio of the entropy changes $\Delta S$ and the magnetization changes $\Delta M$ at the phase transition according to d$B_{\rm C}$/d$T$ = -$\Delta S$/$\Delta M$ (Clausius-Clapeyron relation).~\cite{tari} Applying the experimental results d$B_{\rm C2}$/d$T$\ $\approx 6\cdot 10^{-2}$~T/K and $\Delta M_{\rm C2} = 0.013$~\mb /f.u. yields entropy changes of $\Delta S_{\rm C2} \approx 4$~m\jmk\ associated with \bct , i.e. with changing from AF2 $\rightarrow$ AF3 (see the phase diagram in Fig.~\ref{phd}).

Specific heat data presented in Fig.~\ref{cp} confirm negligible entropy changes at \bct ($T<7$~K). The data show $\lambda$-like anomalies at \tn ($B$) in good agreement to the magnetisation data. The respective values for \tn\ obtained at $B = 0,3,6$ and 9~T (not all data are shown in Fig.~\ref{cp}) are displayed in Fig.~\ref{phd}. In addition to the anomalies at \tn , there is a hump at $T^* \approx 4$~K which is not affected by magnetic fields up to 9~T. It is presumingly not associated to a nuclear Schottky anomaly. At temperatures around \tn , there are clear entropy changes associated with a suppression of long range antiferromagnetic order. In addition, there is a field induced increase of specific heat at $T > 16$~K, i.e. \cp ($B=9$~T) $>$ \cp ($B=0$~T). Quantitatively, \cp ($T=25$~K) increases by about 0.5~\jmk\ upon application of $B=9$~T. This behaviour may suggest the presence of competing ferromagnetic coupling (cf.~\cite{Drechsler07}) which would qualitatively disagree to the DFT results in Ref.~\onlinecite{Johnson13}, i.e. the presence of antiferromagnetic exchange interactions only.

As illustrated in Fig.~\ref{cp}, the phonon contribution is negligible below 5~K. This is illustrated in Fig.~\ref{cp} by dashed lines which show the phonon specific heat for $\Theta_D = 180$, 195, and 210~K, respectively. While \cpp ($\Theta_D=180$~K) exceeds the measured specific heat at 25~K, in the case of $\Theta_D=210$~K the extrapolated anomalous entropy changes exceed the full magnetic entropy so that we conclude $\Theta_D = 195\pm 25$~K. Fig.~\ref{cp}(b) shows the associated magnetic entropy changes \DS ($T$) which have been obtained by integrating $\Delta c_{\rm p}/T$, with \dcp = \cp -\cpp ($\Theta_D = 195$~K). The resulting entropy changes agree to a vanishing field effect at $T \lesssim 7$~K. At $B=0$, nearly 80\% of the whole magnetic entropy changes $R\cdot \ln(2S+1)$ are released below \tn\ which confirms a predominant 3D nature of magnetic order in \mn\ as found by neutron studies and DFT in Refs.~[\onlinecite{Reimers,Johnson13}]. In addition, as mentioned above, the data also imply residual magnetic entropy changes well above \tn .

For $B \leq 9$~T, there is no visible magnetic field effect on \cp\ at $T \lesssim 7$~K. The low-temperature behaviour of \cp\ is approximately proportional to $T^{3/2}$. However, extrapolating the data implies that qualitative changes in the $T$-dependence must be expected for $T<2$~K. The observed low-temperature behavior $c_p \sim T^{n}$ with $n\approx 1.5$ contradicts the magnon specific heat expected in conventional 3D antiferromagnets, i.e. $n=3$ while $n \sim 2$ is the behaviour expected for quasi-2D antiferromagnets. Similar behaviour is often found in frustrated spin systems.~\cite{Ramirez} E.g., the Kagom\'{e}-like jarosite KCr$_3$(OH)$_6$(SO$_4$)$_2$ shows $c_p \sim T^{1.6}$. The data are hence consistent with a frustrated spin system. The small hump in the specific heat data of \mn\ might however suggest an alternative interpretation as it may be associated with coupling of spin and dielectric degrees of freedom. In magnetoelectric \life , changes of the dielectric function below \tn\ are associated with a small hump of the specific heat which is reminiscent of the behaviour observed in Fig.~\ref{cp}.~\cite{DissNeef,LFP}

\section{Electron Spin Resonance}

\begin{figure}[htb]
\includegraphics [width=1\columnwidth,clip] {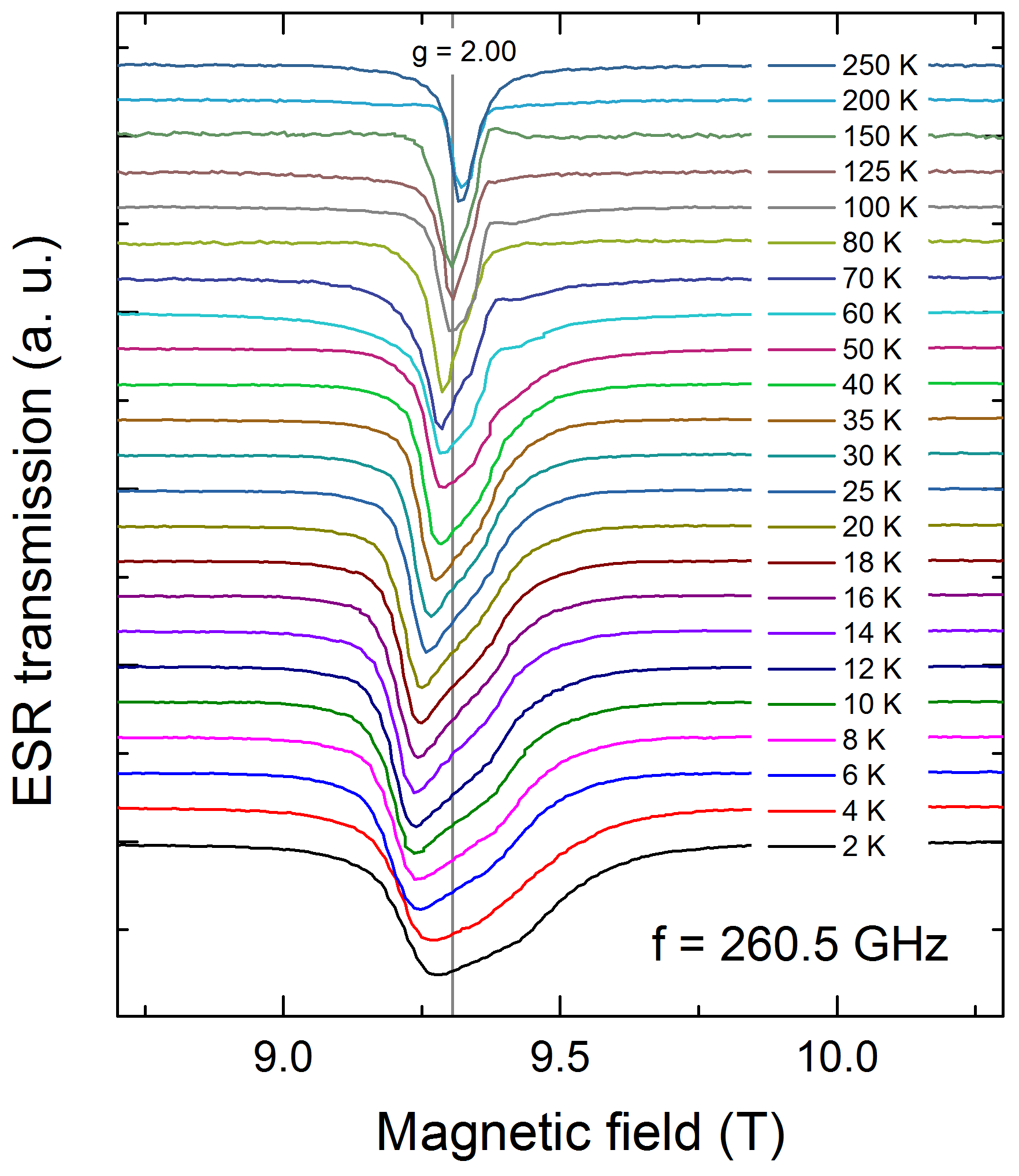}
\caption{Temperature dependence of HF-ESR spectra in the temperature range of $2$ K to $250$ K, at $f = 260.5$ GHz.} \label{TempSpec}
\end{figure}

ESR spectra in Fig.~\ref{TempSpec} taken at $f=260.5$~GHz show a single resonance line at high temperature which shifts and significantly broadens upon cooling. At $T = 250$~K, the data indicate a single ESR line associated with $g = 1.995\pm 0.008$, which is common for high-spin Mn$^{2+}$ in a paramagnetic phase. Upon cooling, the resonance fields shift as temperature approaches \tn , which is typically observed in AFMR spectra due to the evolution of internal fields. At $T=2$~K, i.e. in the long range spin-ordered state where HF-ESR is susceptible to collective magnon modes, a broad and asymmetric resonance feature is found which is typical for powder samples with magnetic anisotropy.

\begin{figure}[htb]
\includegraphics [width=1\columnwidth,clip] {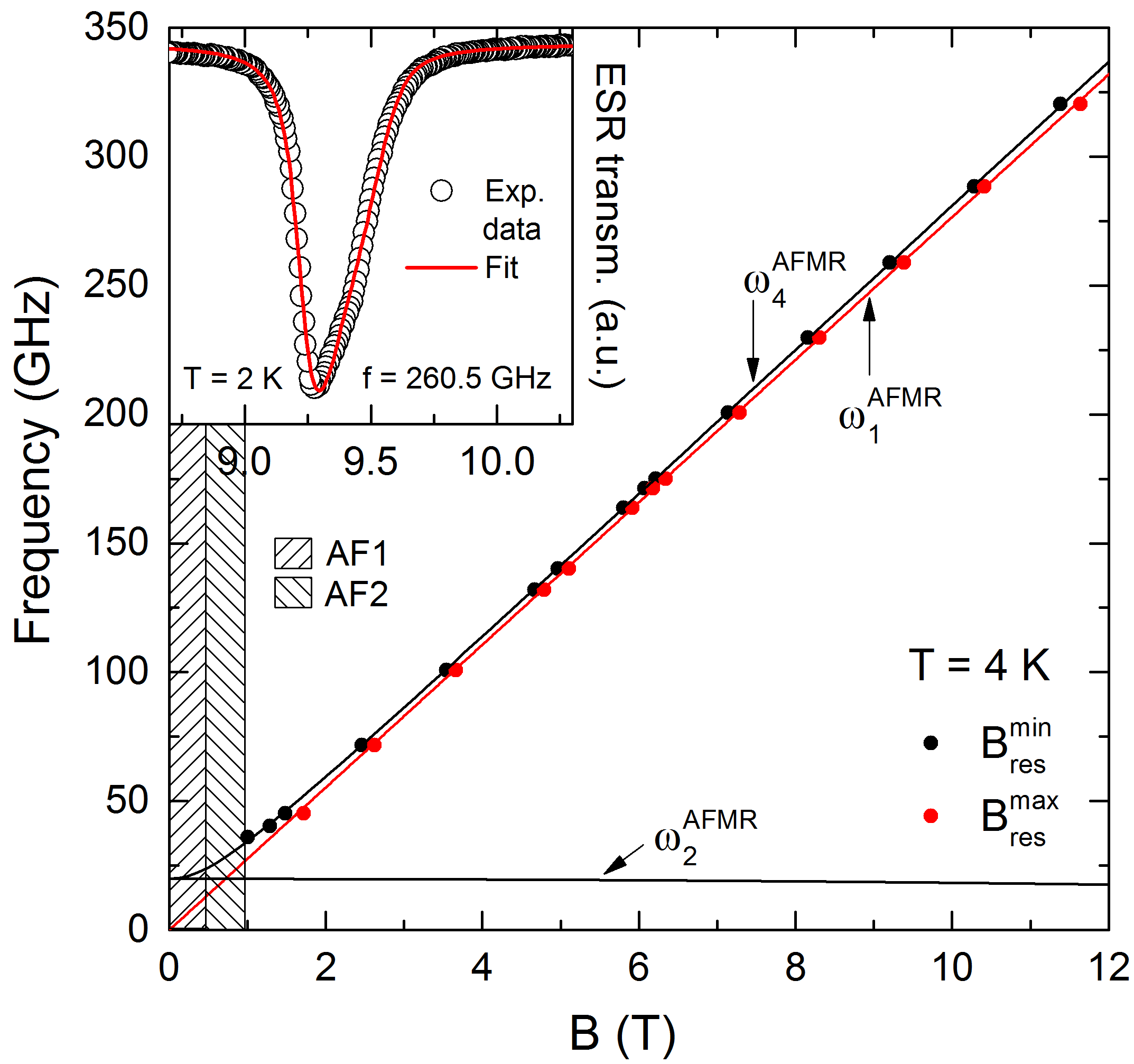}
\caption{Absorption frequencies-field diagram at $T=4$~K. For each frequency, the resonance fields $B_{\text{res}}^{\text{min}}$ and $B_{\text{res}}^{\text{max}}$ have been derived by powder spectra simulations. Solid lines represent fits according to Eqs.~\ref{omega1} to \ref{omega4} (see the text). Hatched areas refer to the phases AF1 and AF2 (cf. Fig.~\ref{phd}). Inset: HF-ESR spectrum at $f = 260.5$~GHz and $T = 2$~K. The line shows the simulated powder spectrum.} \label{spec260}
\end{figure}

Accordingly, the low-temperature spectra are described by means of a powder model which involves different center resonance fields associated with the different orientations of the crystallites with respect to the external field. The simulation yields a good description of the spectra by means of an anisotropic effective resonance field ranging from a minimal resonance center field \bmin\ to a maximal one \bmax\ (see the inset in Fig.~\ref{spec260}).~\cite{easyspin} Note, that broadening due to an inhomogeneous effective field does not describe the spectra well.

\begin{figure}[htb]
\includegraphics [width=1\columnwidth,clip] {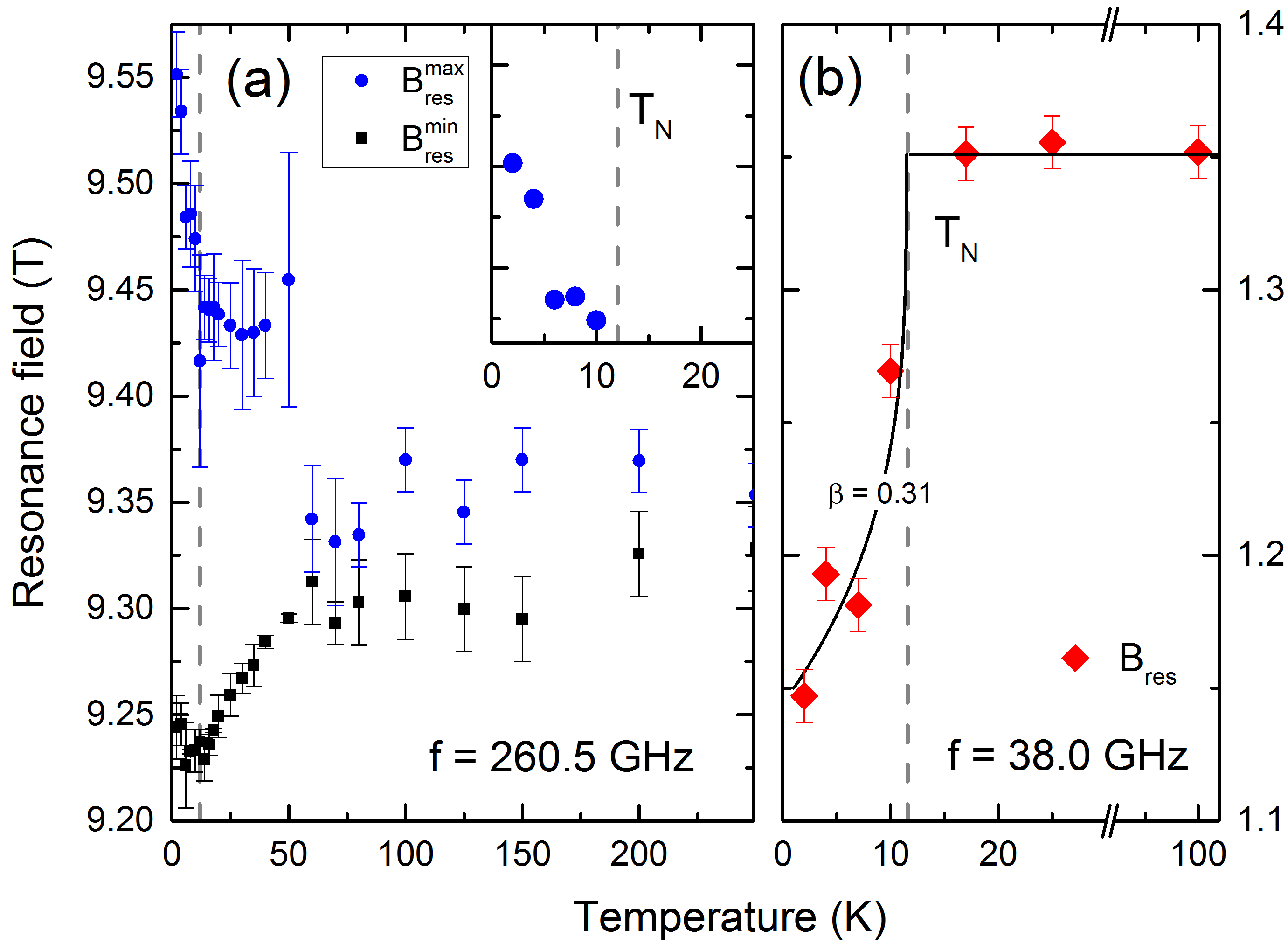}
\caption{(a) Temperature dependence of the HF-ESR resonance fields \bres , at $260.5$ GHz. The resonance fields were obtained from powder spectra simulations. The inset magnifies the behaviour upon crossing \tn . (b) \bres\ vs. $T$ obtained at $38.0$ GHz where only one resonance field is observed. The line is a guide to the eye employing the critical exponent $\beta = 0.31$. Dashed vertical lines show \tn .} \label{TempGval}
\end{figure}

Applying the powder spectra analysis described above allows to study the temperature dependence of the resonance fields \bres\ in more detail. Fig.~\ref{TempGval} shows \bres\ as extracted from the data obtained at $f=260.5$~GHz and $38.0$~GHz, respectively. At $38.0$~GHz, the spectra are described by a single resonance field and there is no significant shift of \bres\ in the paramagnetic phase. A clear shift below \tn\ shows the evolution of an internal magnetic field in the long-range spin ordered phase, i.e. it reflects the magnetic order parameter. The evolution of long range order below \tn\ is also seen at higher frequency $f=260.5$~GHz where the broadened spectra are described by the resonance fields \bmax\ and \bmin , both of which shifting when temperature crosses \tn . In contrast to the low-frequency data, the extracted resonance fields imply the evolution of internal magnetic field well above \tn , i.e. up to $T \sim 60$~K. Note, that the original spectra in Fig.~\ref{TempSpec} indicate inhomogeneous broadening of the resonance even up to 150~K. Both observations imply the presence of internal magnetic field in the paramagnetic phase and hence the evolution of short-ranged magnetic order in this temperature regime. This result agrees to the observed deviation of the static magnetization data from the mean-field description below $\sim 55$~K (cf. Fig.~\ref{chi}) and magnetic entropy changes well above \tn\ (cf. Fig.~\ref{cp}).

Fig.~\ref{spec260} summarizes the frequency dependence of the resonance fields, at $T=4$~K. It comprises the resonance fields \bmax\ and \bmin\ derived by means of the analysis mentioned above. Note, that the resonance feature associated to $\omega$(\bmax ) is less pronounced and is not observed at $f \leq 40$~GHz. For $f \geq 100$~GHz, both resonance branches $\omega(B_{\rm res})$ exhibit a linear field dependence. This linear behavior at high fields implies the $g$-factors $g_{\text{res}}^{\text{min}} = 2.008(3)$ and $g_{\text{res}}^{\text{max}} = 1.982(2)$. Below $\sim 70$~GHz, splitting between the two branches starts to increase and $f(B_{\text{res}}^{\text{min}})$ shows up-turn curvature indicative of finite zero field splitting (ZFS).

A minimal model describing the observed AFMR modes applies a two-lattice system with easy-plane anisotropy. It is motivated by the cycloidal spin structure found in neutron diffraction in which due to large single-ion anisotropy spins are restricted to a plane including the $c$-axis.~\cite{Johnson13} The resulting incommensurate spin order may be approximated by a $60^\circ$ rotation of the spins in the $xy$-plane. Note, that the two-sublattice model agrees to the $M vs. B$ data while three sublattices are usually associated with additional features in the magnetisation curves.

The corresponding minimal Hamiltonian reads:
\begin{align}
	\mathcal{H} =& -\mu_B \sum_j g \vec S_j \!\cdot\! \vec H -\sum_{i j} J_{ij}\vec S_i\!\cdot\!\vec S_j - D^{}\sum_j (\vec S_j^z)^2.
 \end{align}

Here, \mb\ is the Bohr magneton, $g$ the g-value, $D$ the single ion anisotropy, and $J_{ij}$ the exchange coupling between spins $\vec S_i$ and $\vec S_j$. In the mean-field approximation and assuming $B_A \ll B_E$, higher orders of $B_A$ can be neglected. $B_A$ is the anisotropy field and $B_E$ the exchange field. The resulting AFMR modes read:
 \begin{align}
	\omega_1 &= \gamma B  \label{omega1}\\
	\omega_2 &= \gamma\;\sqrt[]{2B_AB_E-(B_A/2B_E)B^2}\label{omega2}
 \end{align}
for magnetic fields $B$ in-plane, and 
 \begin{align}
	\omega_3 &= 0 \label{omega3}\\
	\omega_4 &= \gamma\;\sqrt[]{2B_AB_E-(1-B_A/2B_E)B^2}\label{omega4}
 \end{align}
for magnetic fields $B$ perpendicular to the plane, with $\gamma$ the gyromagnetic ratio. Fitting of the experimental data by means of Eqs.~\ref{omega1} and \ref{omega4} yields an anisotropy of $\sqrt[]{2B_{A}B_{E}}=0.71\pm0.08$~T. The associated zero field splitting amounts to 20~GHz. The fitted resonance branches $\omega_1$ and $\omega_4$ are shown in figure~\ref{spec260}. Extracting the exchange field $B_{E}=B_C/2$ from the saturation field $B_{C}\approx 26$~T (cf. Fig.~\ref{phd}) allows to estimate the planar anisotropy field $B_A=0.02\pm0.01$~T. Note, that this value refers to the two-sublattice model applied here. The resonances associated with the AFMR mode $\omega_2$ are not observed and the shown branch has been calculated from the fitting parameter $B_A$ and from $B_E$. According to the model, it becomes soft well above the field range shown in Fig.~\ref{TempGval}, i.e. at $2B_{E}=26$~T.


\begin{figure}[htb]
\includegraphics [width=0.9\columnwidth,clip] {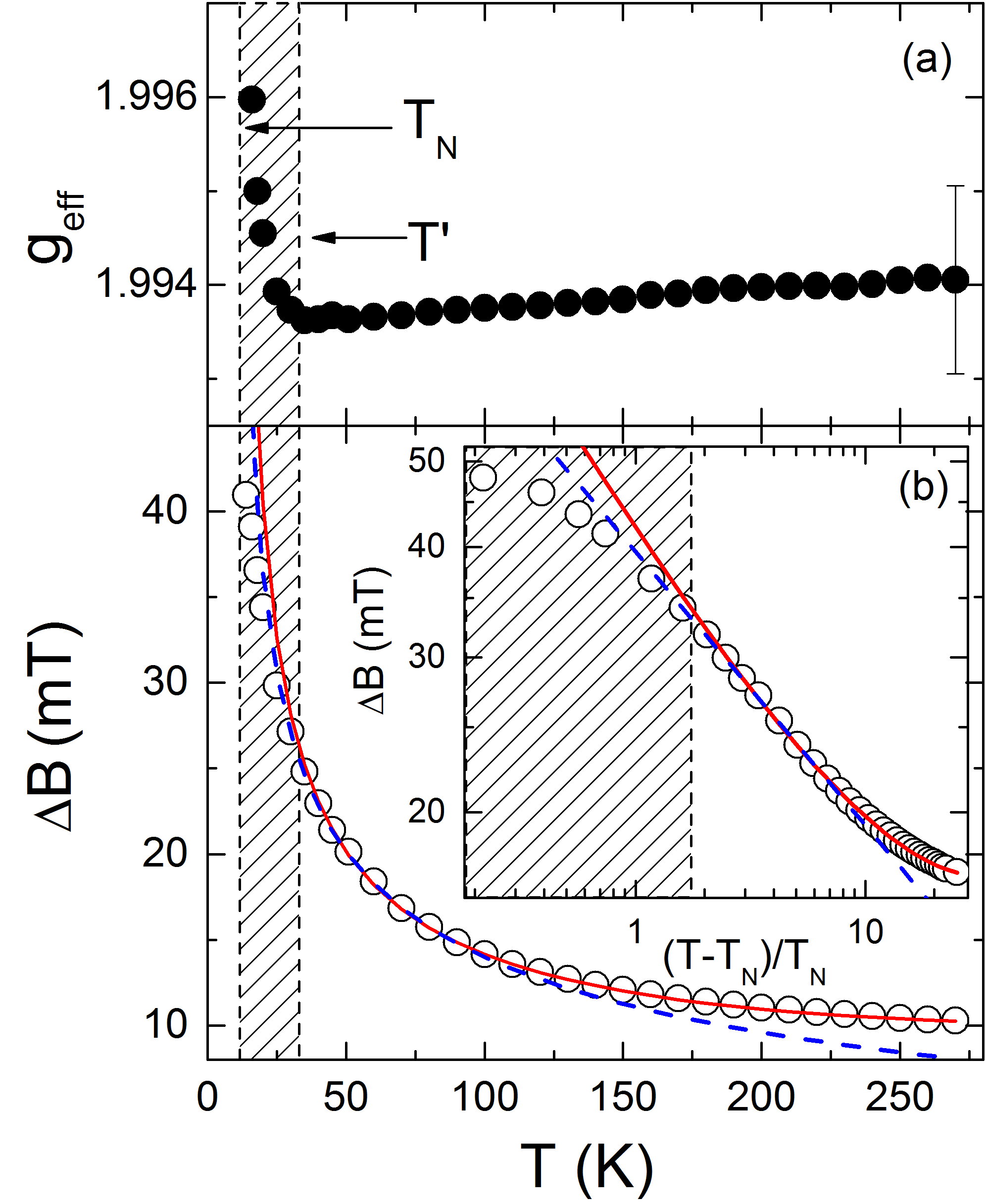}
\caption{(a) Effective $g$-factor \geff\ and (b) linewidth \db\ vs. temperature derived from LF-ESR data. For $T{'}<T<120$~K, \db ($T$) is well described by a power law (see Eq.~\ref{eqesr}) (red and dashed blue line; see the text). The dashed area covers the temperature regime \tn\ $\leq T\leq T{'}$ where the power law fails to describe the data and where \geff\ rapidly increases. At $T>120$~K, the data either imply an additional linear term (red line) or indicate a 2$^{nd}$ crossover regime (see the text). The error bar indicates the systematic uncertainty at room temperature.} \label{lfesr}
\end{figure}

LF-ESR data in the paramagnetic phase show a single exchange-narrowed Lorentzian line which is typical for a concentrated Mn$^{2+}$ spin system. At $T=300$~K in the paramagnetic regime, the average effective $g$-factor amounts to $g = 1.995\pm0.001$ which is in perfect agreement to the HF-ESR data. While \bres\ and hence \geff\ hardly shift upon cooling down to $T{'}\sim 35$~K, there is a pronounced increase of \geff\ at \tn\ $\leq T \leq T{'}$~K, which again signals short-range antiferromagnetic correlations. Consistently with the findings in HF-ESR, the LF-ESR signal vanishes below $T \sim$ \tn\ because of the opening of the spin gap and the line broadening.

Perfect Lorentzian shape of the resonance line at 300~K rules out spin diffusion. Upon cooling, the linewidth \db\ continuously increases which may be due to critical behaviour associated to the evolution of spin-spin correlations. The results are hence consistent with the presence of high-temperature antiferromagnetic correlations. Linewidth broadening can be described in terms of a power law of the reduced temperature, with the critical exponent $p$ being associated with the spin dimensionality and the magnetic anisotropy of the system. With the temperature independent linewidth $\Delta B(\infty)$ at infinite temperature, the critical slowing down is analyzed as follows:

\begin{equation}\Delta B(T) - \Delta B(\infty) \propto A\cdot \left(\frac{T-T_{\rm N}}{T_{\rm N}}\right)^{-p}+B\cdot(T-T_{\rm N}). \label{eqesr} \end{equation}

The linear term ($B>0$) is suggested by the data at $T\gtrsim 120$~K (see the inset of Fig.~\ref{lfesr}). Eq.~\ref{eqesr} appropriately describes \db ($T$) above $T\simeq T'$, with $p = 0.47(2)$ and $\Delta B(\infty)\approx 2$~mT (red line). Excluding the linear term yields $p = 0.32(1)$ (dashed blue line). Upon approaching \tn , non-linear behavior in the log-plot implies failure of a power law description. In terms of Eq.~\ref{eqesr}, the curved function would result in a continuous change of $p$ below $T'$. The critical divergence becomes weaker upon approaching the 3D cycloidic spin ordered phase.

The observed critical exponent is much smaller than what is theoretically expected for 2D and 3D antiferromagnets~\cite{Benner1990} or what is found in 1D antiferromagnets.~\cite{Arts1983,Ajiro1975} In contrast, it is similar to the bahaviour in the easy-plane magnet CsMnF$_3$ ($p = 0.51$)~[\onlinecite{Witt1972}], in the 2D quantum antiferromagnet SrCu$_2$(BO$_3$)$_2$ ($p=0.51$)~[\onlinecite{Zorko}], or in the 2D Ising antiferromagnet MnTiO$_3$ ($p=0.49$) which is a linear magnetoelectric~[\onlinecite{Akimitsu,Mufti}]. We also note similar behaviour in triangular systems which are discussed in terms of the vicinity to a Berezinskii-Kosterlitz-Thouless (BKT) phase.~\cite{Ajiro} In general, the critical behaviour found in \mn\ is consistent with findings in systems with competing interactions.

Spin-lattice interaction may account for a (positive) linear-$T$ contribution to \db\ as proposed for manganites in Ref.~[\onlinecite{Huber2014}]. In these systems, an exchange narrowed line is considered where phonon modes with frequencies below a spin-spin relaxation rate contribute to the linewidth. Note, however, that a further change of the critical behaviour towards $p\approx 0.17$ would describe the high-$T$ behaviour without employing an additional linear term. The strong increase of ESR linewidth with temperature found in \cu\ can be ascribed to thermal activation of a dynamic Jahn-Teller effect\cite{Heinrich} which is not present in the Mn$^{2+}$-system at hand.

\section{Discussion}

The HF-EPR data imply a small but finite planar anisotropy showing up in the zero-field splitting of the associated AFMR mode of approximately 20~GHz. Our analysis by means of an easy-plane two-sublattice model yields the anisotropy field $B_A=0.02$~T which is in the typical range for Mn$^{2+}$-ions. E.g., in CaMnCl$_3\cdot$H$_2$O, $B_A$ amounts to 0.06~T.~\cite{Phaff} Our finding of the planar-type anisotropy qualitatively confirms the DFT results in Ref.~[\onlinecite{Johnson13}]. Based on these band structure calculations, Johnson \etal\ emphasize a crucial role of anisotropy for stabilizing the cycloidal spin structure in \mn\ as it favors cycloidic over helical spin structure in the ground state. Our experimental data presented here however imply only very small anisotropy.

The presence of only small magnetic anisotropy is corroborated by the magnetic phase diagram which exhibits field induced phase transitions at low magnetic fields only. However, $two$ field induced phase transitions are observed at \bco\ $\approx$ 0.5~T and \bct\ = 0.9~T, respectively, at $T=2$~K. Typically, small anisotropy yields conventional spin-flop transitions at moderate magnetic field as illustrated in CaMnCl$_3\cdot$H$_2$O at $B_{\rm sf} = 1.6$~T.~\cite{Phaff} Our analysis of the slope of the phase boundary (Fig.~\ref{phd}) indicates only small entropy differences between the associated antiferromagnetic phases AF2 and AF3. Following Ref.~[\onlinecite{Johnson13}] one may speculate whether external magnetic fields affects the subtle interplay between the cycloidal spin structure, i.e. the ground state, and the helical one by tipping the energy balance towards the latter at \bco . While, \bct\ may be associated to spin reorientation. The example of \bacu\ illustrates the possibility of a two-stage spin reorientation phenomenon. In this system, the relative arrangement of adjacent long range antiferromagnetically ordered spins is hardly affected by external magnetic fields while the whole spin structure is reoriented in two phase transitions.~\cite{Zhel} The AF3 phase probed by our HF-ESR and magnetisation data at $B \gtrsim 1$~T is neither described with a three-sublattice model nor with six sublattices if magnetic coupling of magnitude reported in Ref.~[\onlinecite{Johnson13}] for cycloidic order in AF1 is used. The fact that, in AF1 and AF2, no resonance is found, at $\geq 35$~GHz and at $B < 1$~T, suggests that AF1 and AF2 exhibit only small ZFS and hence small anisotropy, too, similar to the findings in AF3. A possible larger ZFS would show up in the HF-ESR data at $B<1$~T and higher frequency. We however emphasize that the magnetic spectroscopies applied here do not discriminate between different spin structures as long as the associated AFMR modes can be described by means of a two-sublattice model. In addition, one may speculate whether the unusual temperature dependence of the specific heat, i.e. a field-independent hump-like region in the ordered phase, is associated with the interplay of anisotropy and energetically neighboring states, too. Very recently, the ground state at $B=0$ was found to exhibit a small tilting from the (110) plane in the spin-spiral plane.~\cite{Tokura2016} It is speculated that this tilting could arise from a small angle between the direction of the anisotropy and the [001]-axis. We note, that the HF-ESR data presented here probe the magnetic phase above $1$~T.

Finally, our results imply short-range magnetic correlations well above \tn\ as consistently indicated by the shift of HF-ESR resonance fields below $\sim 60$~K, the failure of a power law description in the LF-ESR data, and the observation of magnetic entropy changes at temperatures well above \tn . Such a wide range of antiferromagnetic fluctuations agrees to the structurally triangular, $viz.$ magnetically frustrated arrangement of Mn$^{2+}$-ions in \mn . The failure to describe the LF-ESR linewidth in terms of a single power-law may indicate dimensional crossover when approaching the chiral magnetic ground state. \\

\section{Summary}

To summarize, we report the static and dynamic magnetic properties as well as specific heat of structurally and magnetically chiral \mn . The magnetic phase diagram shows competing antiferromagnetic phases at small magnetic fields, i.e. at \bco\ $\approx$ 0.5~T and \bct\ = 0.9~T. The AFMR modes imply planar anisotropy which qualitatively confirms recent predictions.~\cite{Johnson13} The data are well described by means of an easy-plane two-sublattice model with the anisotropy field $B_A=0.02$~T. The exchange field $2B_{\rm E}= 26$~T is obtained from the saturation field of the static magnetization. The results are discussed in terms of a delicate balance of cycloidal and helical spin structures under the influence of small planar anisotropy and external magnetic field. The frustrated nature of the system is reflected by the fact that short-range magnetic correlations are observed well above \tn .\\

\begin{acknowledgements}
We are very grateful to V.B. Nalbandyan and A.Yu. Nikulin for providing the sample for this study. R.K. and E.A.Z. acknowledge financial support by the Excellence Initiative of the German Federal Government and States. E.A.Z. appreciates support from the Russian Foundation for Basic Research (grants 14-02-00245). This work was supported in part from the Ministry of Education and Science of the Russian Federation  in the framework of Increase Competitiveness Program of NUST "MISiS" (No. K2-2015-075) and by Act 211 Government of the Russian Federation, contract No. 02.A03.21.0006.

\end{acknowledgements}

\end{document}